\documentclass{elsart5p}

\usepackage{graphicx}
\usepackage{bm} 
\usepackage{epsfig}
\usepackage{bm}

\def\vec#1{\mathchoice{\mbox{\boldmath$\displaystyle#1$}}
{\mbox{\boldmath$\textstyle#1$}}
{\mbox{\boldmath$\scriptstyle#1$}}
{\mbox{\boldmath$\scriptscriptstyle#1$}}}
\makeatletter
\newcommand\erfc{\mathop{\operator@font erfc}\nolimits}
\def\slashchar#1{\setbox0=\hbox{$#1$}
   \dimen0=\wd0 \setbox1=\hbox{/} \dimen1=\wd1
   \ifdim\dimen0>\dimen1 \rlap{\hbox to \dimen0{\hfil/\hfil}} #1
   \else  \rlap{\hbox to \dimen1{\hfil$#1$\hfil}} / \fi}

\newcommand{\NP}{{\rm NP}}

\begin{document}

\begin{frontmatter}

\title{Correlations between perturbation theory and power corrections
  \\ in QCD at zero and finite temperature } \thanks{Supported by
  Spanish DGI and FEDER funds with grant FIS2008-01143, Junta de
  Andaluc{\'\i}a grant FQM-225-05, EU Integrated Infrastructure
  Initiative Hadron Physics Project contract
  RII3-CT-2004-506078. E.~Meg{\'\i}as is supported by the Alexander
  von Humboldt Foundation.}

\author[ITP]{E. Meg{\'\i}as},
\ead{emegias@tphys.uni-heidelberg.de}
\author[FAMN]{E. Ruiz Arriola},
\ead{earriola@ugr.es}
\author[FAMN]{L. L. Salcedo}
\ead{salcedo@ugr.es}

\address[ITP]{
Institute for Theoretical Physics,
University of Heidelberg,
D-69120 Heidelberg, Germany
}
\address[FAMN]{
Departamento de F{\'\i}sica At\'omica, Molecular y Nuclear,
Universidad de Granada,
E-18071 Granada, Spain
}


\date{\today}

\begin{abstract}
  The duality between QCD perturbative series and power corrections
  recently conjectured by Narison and Zakharov is analyzed. We propose to
  study correlations between both contributions as diagnostics tool. A
  very strong correlation between perturbative and non perturbative
  contributions is observed for several observables at zero and at
  finite temperature supporting the validity of the dual description.
\end{abstract}


\end{frontmatter}

\section{Introduction}
\label{sec:intro}

The disentanglement between perturbative and non perturbative effects
in QCD has been a major enterprise in the last three decades since the
sum rules technique was first
suggested~\cite{Shifman:1978bx,Shifman:1978by}. While short distance
radiative corrections are perturbatively computable and are
characterized by a smooth logarithmic dependence on the relevant
energy scale, non perturbative effects manifest as a stronger
power-like dependence and in terms of vacuum expectation values of
local and low dimensional gauge invariant operators. Although the
underlying quark-gluon dynamics should determine the relative strength
of perturbative and non perturbative contributions unambiguously, up
till now these condensates have been treated {\it de facto} as
independent parameters, unrelated to the first few terms of the
perturbative series. During many years renormalons have been viewed as
a bridge between perturbative and non perturbative physics~(see e.g.
Ref.~\cite{Zakharov:1997xs,Zakharov:2003vd} and references therein for
a review). In a recent paper Narison and
Zakharov~\cite{Narison:2009ag} have conjectured a quite different
scenario, namely a duality between condensates and perturbative
contributions. This duality concerns the properties of large order
perturbative series which involves an expansion in the strong running
coupling constant $\alpha_s(Q^2)$, and it establishes that they are
dual to non perturbative power corrections, i.e. powers of
$(\Lambda_{\rm QCD}/Q)^n$. At the practical level, this means that if
one considers a short perturbative series then one should add the
leading power correction by hand. Only when one uses long perturbative
series there is no reason to add power corrections. The confirmation
of this conjecture might pave the way for practical approaches where
condensates are required; the need of allowed and forbidden local
condensates would be justified because of a lack of a complete
perturbative series to all orders. This is quite timely since in many
applications at zero temperature and finite temperature the
phenomenological need for power corrections involving dimension-2
operators is overwhelming. The zero temperature example {\it par
  excellence} is given by the heavy $q\bar{q}$ potential, where the
string tension, a dimension-2 object not related to any local gauge
invariant operator, appears as the unequivocal signal of confinement.
There have been speculations on the appearance of dimension 2
contributions to the average plaquette~\cite{Burgio:1997hc}. Recent
lattice calculations~\cite{Ilgenfritz:2009ck} computing the average
plaquette in lattice field theory up to 20-th order do not yet see the
onset of renormalon physics; a mild geometric type perturbative series
which successfully reconstructs the full result, is observed instead.

While high momenta at zero temperature probes the theory in the
asymptotically free region, we note that a parallel discussion for
finite temperature $T \gg \Lambda_{\rm QCD} /2 \pi $ above the
deconfinement phase transition might be carried out. However, the
explicit breaking of Lorentz invariance triggered by the privileged
heat bath reference frame makes the theoretical discussion much more
involved~\cite{Kraemmer:2003gd,Shuryak:2008eq}. At finite temperature
the recent discovery of inverse temperature power corrections from
relatively old lattice data becomes evident from plots in $1/T^2$
and has been quite impressive and rather unexpected.  They can
effectively be explained by a dimension two condensate, starting from
the Polyakov loop~\cite{Megias:2005ve,Megias:2006ke}, the heavy
quark-antiquark free energy in~\cite{Megias:2007pq} as well as the
trace anomaly and QCD equation of state
in~\cite{Megias:2008dv,Megias:2008rm,Megias:2009mp} (see also
Ref.~\cite{Pisarski:2006yk}). Of course, it would be quite interesting
to determine whether these thermal power corrections are dual, in the sense
of Narison and Zakharov, to a long perturbative series. The present
paper addresses this important issue.

While the duality conjecture might eventually be tested more
convincingly in the future, we suggest an alternative approach where
some quantitative insight may also be gathered both at zero and finite
temperature from confronting current lattice and perturbative
results. Basically, the idea is quite simple. Given that the only
scale entering the perturbative series to any order is $\Lambda_{\rm
  QCD}$, in a fit to lattice data containing {\it both} perturbation
theory and condensates we should observe 1) smaller contributions from
condensate at increasing perturbative orders and 2) a rather strong
statistical correlation between $\Lambda_{\rm QCD}$ and the
dimensionful condensates. 

In the present paper we pursue this idea to check with the accessible
theoretical information and lattice data the validity of the duality
conjecture.  In Sec.~\ref{sec:qqpot} we deal first with the more
familiar quark-antiquark potential at zero temperature, where we
indeed observe, within uncertainties, the expected correlations. This
not only supports the perturbative-power duality but also qualifies
the correlation method as a handy tool to study the duality elsewhere.
We are interested to do so at finite temperature above the
deconfinement phase transition for the Polyakov loop in
Section~\ref{sec:Polyakov_loop} and the trace anomaly in
Section~\ref{sec:trace_anomaly}. As a useful guideline we use the
finite temperature model where non perturbative thermal power
corrections are driven by a dimension 2 gluon condensate in the
dimensionally reduced
theory~\cite{Megias:2005ve,Megias:2006ke,Megias:2007pq,Megias:2008dv,Megias:2008rm,Megias:2009mp}.
While the model might be improved, taken at face value it provides a
unified and coherent description of gluodynamics lattice data using
the {\it same} values of condensates within estimated errors. It
therefore provides an ideal playground to search for possible
correlations in the sense of the above mentioned duality.



\section{The quark-antiquark potential}
\label{sec:qqpot}

Our points are best exemplified with the (zero temperature) heavy
quark-antiquark potential within a perturbative expansion for which
many efforts have been devoted. Nowadays the perturbative series are
partially known up to order $\alpha_s^4 \log^2 \alpha_s$ in the Weyl
or temporal gauge, $A_0=0$ (see Ref.~\cite{Brambilla:2009bi} and
references therein\footnote{See also
  \cite{Anzai:2009tm,Smirnov:2009fh} for a recent complete three loop
  calculations in Feynman gauge.}). In either case it is known that
this perturbative computation can only reproduce lattice data at small
separations, and fails for $r > 0.25\,{\rm fm}$. The Cornell potential
is a phenomenological version of this potential and it gives a good
overall description of the lattice data~\cite{Necco:2001xg} for all
separations. It reads
\begin{equation}
V_{q\bar{q}} (r) = -\frac{4}{3} \frac{\alpha_s}{r} + \sigma  r \,, 
\label{eq:cornell}
\end{equation} 
where $\alpha_s$ is the QCD coupling constant, which is considered as
a constant, i.e. no running, and $\sigma \simeq 4.64\, {\rm fm}^{-2}$
is the string tension term. The Coulomb term corresponds to the
leading order in perturbation theory. The linear term follows from
quarkonium phenomenology, and it is widely accepted that it cannot
follow from a perturbative computation. This leads to the conclusion
that the perturbative potential is not complete, and should be
extended with a linear term put by hand, i.e.
\begin{equation}
V_{q\bar{q}}(r) = \frac{1}{r} \sum_{n=1}^N a_n \alpha^n_s(r) + \sigma_N r \,. 
\label{eq:Vqqpert}
\end{equation}  
Because the accepted separation between perturbative and non
perturbative contributions, one is tempted to identify the parameters
$\sigma_N$ and $\sigma$. In this section it will be shown that this
identification might not be correct, in line with
Ref.~\cite{Narison:2009ag}, and in fact there is a mixing between
power-like corrections and the perturbative series. This means that
eventually $\lim_{N \to \infty } \sigma_N = 0$. In order to provide
further convincing evidence that this might happen, we will analyze
lattice data for the heavy $q\bar{q}$ potential from
Ref.~\cite{Necco:2001xg} using Eqs.~(\ref{eq:cornell}) and
(\ref{eq:Vqqpert}).

Two kinds of fits are considered. In the first ones, a distance
interval is sought where the perturbative expansion works well. As
expected, this corresponds to taking a sufficiently small distance
region, namely, $ 0.085\,\textrm{fm} < r < 0.170\,\textrm{fm}$.  This
regime follows from the requirement that the fitted $\sigma_N$ is
compatible with zero within errors.  The second type of fits use all
distances of lattice data, $0.085\,\textrm{fm} < r <
0.830\,\textrm{fm}$.  In both cases an additive constant is allowed in
the potential, chosen so that the fit reproduces exactly a point of
lattice data at an intermediate distance $r=0.25 \,{\rm fm}$.  It has
been checked that the conclusions are unchanged when this parameter is
also included in the fit.

\begin{table}[htb]
\begin{center}
\begin{tabular}{|c|c|c|c|c|}
\hline
{\bf Order}  &  $\delta$  &    $\sigma [{\rm fm}^{-2}]$   &   $r(\sigma,\delta)$
 & $\chi^2/{\rm dof}$  \\
\hline
Tree Level                                 & $-$  & 4.99(11) &  0.991  &  0.98  \\
1-loop                                     & 3.66(7)  & 4.25(6)  &  0.974  &  0.20  \\
2-loop                                     & 4.54(10) & 4.12(6)  &  0.978  &  0.79  \\
N$^3$LO~\cite{Anzai:2009tm,Smirnov:2009fh} & 4.31(13) & 4.07(6)  &  0.980  &  0.93  \\
N$^3$LL~\cite{Brambilla:2009bi}            & 4.19(14) & 3.85(7)  &  0.984  &  0.76   \\
\hline
\end{tabular}
\end{center}
\caption{
Fit, using Eq.~(\ref{eq:Vqqpert}), of heavy $q\bar{q}$ potential
lattice data from Ref.~\cite{Necco:2001xg}.  The fit is performed in
the interval $0.085 \,{\rm fm} < r < 0.332 \,{\rm fm}$ for the tree
level, and in the full interval of lattice data, $0.085 \,{\rm fm} < r <
0.830 \,{\rm fm}$, for the rest.  The intervals are chosen so that
$\chi^2/{\rm dof} < 1$. At three level, the correlation coefficient
refers to $-r(\sigma,\alpha_s)$.
}
\label{tab:Vqqall}
\end{table}

\begin{figure}[tbp]
\begin{center}
\epsfig{figure=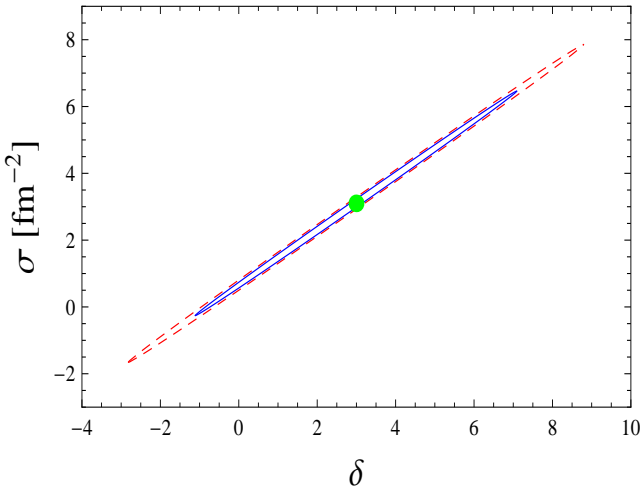,height=6.5cm,width=8.5cm}
\end{center}
\caption{Correlation ellipses corresponding to $\Delta\chi^2 = 2.3,
  4.6 $, between the perturbative parameter $\delta =
  \log(\mu/\Lambda_{\rm QCD})$ and the non perturbative $\sigma$ for
  the heavy $q\bar{q}$ potential. The perturbative formula has been
  considered at 3 loops order (N$^3$LL)~\cite{Brambilla:2009bi}, i.e.
  $N=4$ in Eq.~(\ref{eq:Vqqpert}). The fit uses data in the interval
  $0.085 \,{\rm fm} < r < 0.170 \,{\rm fm}$.}
\label{fig:Vqq1}
\end{figure}

\begin{figure}[tbp]
\begin{center}
\epsfig{figure=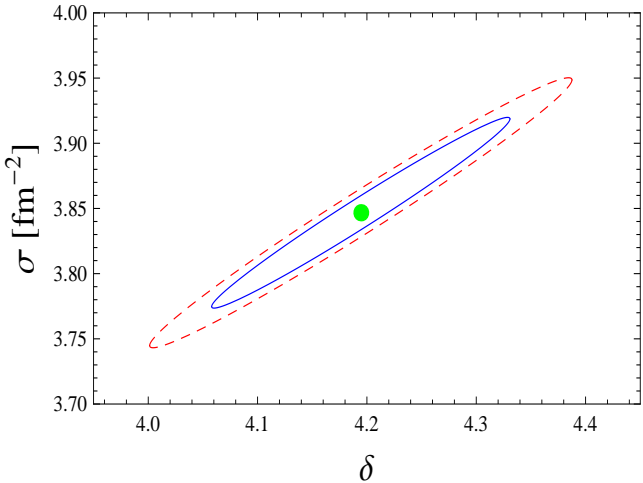,height=6.5cm,width=8.5cm}
\end{center}
\caption{Same as Fig.~\ref{fig:Vqq1}, but for the last line in Table
  ~\ref{tab:Vqqall}. The fit uses data in the interval $0.085 \,{\rm
    fm} < r < 0.830\, {\rm fm}$ }
\label{fig:Vqq2}
\end{figure}

For the small distance regime, the perturbative series to ${\cal
  O}(\alpha_s^4)$ describes very well by itself the lattice data
($\chi^2 /{\rm dof} = 0.04 \ll 1$) and the string tension $\sigma_N$
turns out to be compatible with zero.  The fits including all lattice
data are summarized in Table~\ref{tab:Vqqall}. The value of $\sigma_N$
tends to decrease (within errors) as higher orders in perturbation
theory are included, and this is what would be expected from the
scheme of duality between power corrections and perturbation theory.

The most relevant feature uncovered by this analysis is the strong
correlation found between the perturbative series and the string
tension term, not only in the small distance regime, but also in the
entire distance regime. Also the correlation is larger at higher
orders.\footnote{A fit in the interval $0.085 \,{\rm fm} < r < 0.830
  \,{\rm fm}$ using the tree level leads to
  $r(\sigma,\alpha_s)=-0.960$.} Figs.~\ref{fig:Vqq1} and
\ref{fig:Vqq2} show the correlation ellipses when terms up to N$^3$LL
are included in the potential. This strong correlation confirms the
dual description of QCD proposed by Narison and
Zakharov~\cite{Narison:2009ag}. In what follows we aim to
apply the same correlation method to determine whether or not this
duality takes place also at finite temperature above the deconfinement
phase transition.

\section{The Polyakov loop}
\label{sec:Polyakov_loop}

The vacuum expectation value of the Polyakov loop in a gauge in which
$A_0$ is time independent reads
\begin{equation}
L(T) = \left\langle \frac{1}{N_c} \, {\rm tr}_c \, e^{i g A_0({\mathbf x})/T } 
\right\rangle 
\,,
\label{eq:defPL}
\end{equation}
and an expansion of the exponential  gives
\cite{Megias:2005ve}:
\begin{equation}
 \log L(T) =  - \frac{g^2 
\langle A_{0,a}^2\rangle}{4 N_c T^2}  + {\cal O}(g^5) \,.
\label{eq:LA2A4}
\end{equation}
It has been shown in a series of
works~\cite{Megias:2005ve,Megias:2006ke,Megias:2007pq,Megias:2008dv,Megias:2008rm,Megias:2009mp}
that power corrections provide the bulk of observables at finite
temperature in the non perturbative regime of the deconfined phase of
QCD, i.e. in the regime $ T_c < T < 6 T_c$.  Our considerations were
first proposed to describe the lattice data for the renormalized
Polyakov loop in this regime, and follows from the introduction of a
tachyonic gluon mass at short distances in the gluon
propagator~\cite{Megias:2005ve}. This is the analog of the zero
temperature modification proposed in Ref.~\cite{Chetyrkin:1998yr}.
Moreover, as shown in \cite{Megias:2007pq}, a common value for this
mass reproduces both the string tension and Polyakov loop data. These
considerations imply that the perturbative value of $\langle
A_{0,a}^2\rangle$ should be augmented with a non perturbative term
directly related to the tachyonic mass~\cite{Megias:2005ve}:
\begin{equation}
  \langle A_{0,a}^2\rangle = 
\langle A_{0,a}^2\rangle^{\rm P} + \langle A_{0,a}^2\rangle^{\rm NP}_{T} \,.
\label{eq:A0pA0np}
\end{equation}
Up to radiative corrections, the perturbative part $\langle
A_{0,a}^2\rangle^{\rm P}$ is proportional to $T^2$ whereas $\langle
A_{0,a}^2\rangle^{\rm NP}$ is temperature independent. Thus, the total
Polyakov loop can be separated into perturbative and non perturbative
contributions in $\langle A_{0,a}^2\rangle$, and reads ($N_c=3$)
\begin{equation}
\log L(T) = \log L_{\rm P}(T) 
-\frac{g^2 \langle A_{0,a}^2\rangle^{\rm NP}}{12 T^2} \,.
\label{eq:PLpnp}
\end{equation}
The presently available perturbative calculations have been carried
out to order $g^4$ \cite{Gava:1981qd} (recently corrected in
\cite{Burnier:2009bk}).\footnote{ For gluodynamics with
  $N_c=3$ this gives
\begin{equation}
\log L_{\rm P}(T) =  \frac{g^3}{6\pi}
+ \frac{g^4}{4\pi^2}\left(\log g + \frac{1}{4}\right)
+ {\cal O}(g^5)  
\,,
\label{eq:LBurnier}
\end{equation}
(where the subindex P stands for perturbative).  Since the $\beta$
function starts at order $g^3(\mu)$, changes in $\mu$ affect ${\cal
  O}(g^5) $.  } Clearly, Eq.~(\ref{eq:PLpnp}), with a perturbative
series plus a dimension 2 power-like term, resembles
Eq.~(\ref{eq:Vqqpert}).

\begin{figure}[tbp]
\begin{center}
\epsfig{figure=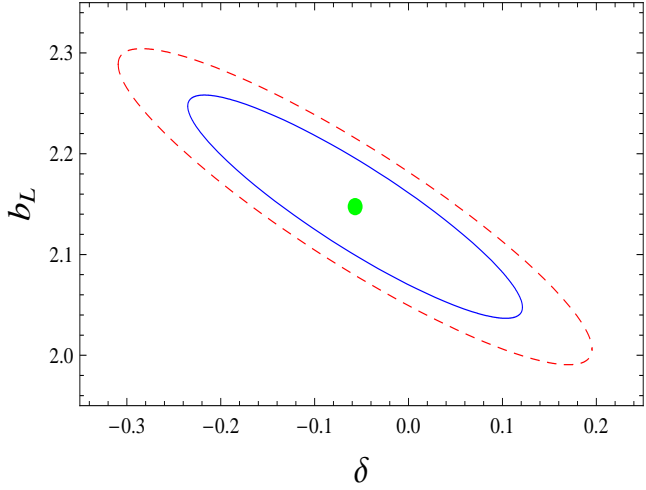,height=6.5cm,width=8.5cm}
\end{center}
\caption{Correlation ellipses for the Polyakov loop corresponding to
  $\Delta\chi^2 = 2.3, 4.6$ between the perturbative
  parameter $\delta$ and the non perturbative parameter $b_L$. The
  perturbative Polyakov loop is included to order $g^4$.}
\label{fig:PL1}
\end{figure}

The lattice data from Ref.~\cite{Gupta:2007ax} for the renormalized
Polyakov loop with $N_c=3$ can be fitted using Eq.~(\ref{eq:PLpnp}),
including the non perturbative term and different orders in the
perturbative series. The scale in $g(\mu)$ is taken as $\mu=e^\delta
2\pi T$.  Results are shown in Table~\ref{tab:PL1}. Within
uncertainties it can be seen that the coefficient of the power-like term
$b_L=g^2 \langle A_{0,a}^2\rangle^{\rm NP}/6 T_c^2$ does not change
much while the correlations are quite strong. The corresponding
correlation ellipses are displayed in Fig.~\ref{fig:PL1}.
\begin{table}[htb]
\begin{center}
\begin{tabular}{|c|c|c|c|c|c|}
\hline
{\bf Order}  &     $L_{\rm P}(6 T_c)$ & $\delta$  &    $b_L$   &   $r(b_L,\delta)$ & $\chi^2/{\rm dof}$ \\
\hline
$L_{\rm P} = {\rm const}$   &  1.121(8)   & $-$  & 1.72(5)  & $-0.472$   & 0.45    \\
${\cal O}(\alpha^{3/2})$  &  1.125(11)   & $-0.72(20)$ & 2.23(16) & $-0.957$  & 1.22   \\
${\cal O}(\alpha^2)  $      &  1.123(9)   & $-0.06(18)$ & 2.15(11)  &  $-0.901$  &  1.44   \\
\hline
\end{tabular}
\end{center}
\caption{Fit of lattice data for the renormalized Polyakov loop from
  Ref.~\cite{Gupta:2007ax}, for $N_\sigma^3\times N_\tau = 32^3 \times
  8$, using Eq.~(\ref{eq:PLpnp}) in the interval $ 1.03 < T/T_c < 6$.  The
  fit is made using the perturbative series $L_{\rm P}(T)$ up to a given
  order, c.f.  Eq.~(\ref{eq:LBurnier}), in addition to the non
  perturbative term of Eq.~(\ref{eq:PLpnp}).  $b_L$ stands for $g^2
  \langle A_{0,a}^2\rangle^{\rm NP}/6 T_c^2$. In the first line 
  $L_{\rm P}$ is taken to be a constant and the correlation coefficient
  there refers to $-r(b_L,L_{\rm P})$.}
\label{tab:PL1}
\end{table}

The possible contribution of a dimension four condensate in $L(T)$ can
also be considered. To this end, we carry out a fit with the formula:
\begin{equation}
  \log L(T) = \log L_{\rm P}(T) -\frac{b_L}{2} \left(\frac{T_c}{T}\right)^2 
- \frac{c_L}{2} \left(\frac{T_c}{T}\right)^4  
\,.
\label{eq:PLpnp4}
\end{equation}
The results are shown in Table~\ref{tab:PL2}. It is noteworthy that
the correlations $r(b_L,\delta)$ tend to increase with the
perturbative order. Although the fit does not yield a clear signal for
$c_L$, the correlation $r(b_L,c_L)$ is remarkably high. This is
in line with some appreciation in~\cite{Narison:2009ag} about the
eventual extension of the duality also to dimension four condensates.

\begin{table}[htb]
\begin{center}
\begin{tabular}{|c|c|c|c|c|c|c|c|}
  \hline
  {\bf Order}  &  $L_{\rm P}(6T_c)$ & $\delta$  &    $b_L$  &  $c_L$  &   $r(b_L,\delta)$ & $r(b_L,c_L)$ & $\frac{\chi^2}{\rm dof}$ \\
  \hline
  $ {\rm const}L_{\rm P} $   & 1.118(12)  & $-$  & 1.61(27) & 0.13(34)  & $-0.605$ & $-0.974$   & 0.41    \\
  ${\cal O}(\alpha^{3/2})$  & 1.128(16)  & $-0.78(28)$   & 2.39(46) & $-0.14(33)$ & $-0.845$ & $-0.871$  & 1.55   \\
  ${\cal O}(\alpha^2)$                          & 1.128(15) & $-0.15(27)$ & 2.36(44)  &  $-0.21(37)$  & $-0.828$ & $-0.940$ &  1.61   \\
\hline
\end{tabular}
\end{center}
\caption{Same as table~\ref{tab:PL1} but including a term~$\sim
  c_L/T^4 $, c.f. Eq.~(\ref{eq:PLpnp4}). The fit is made in the interval $ 1.03 <
  T/T_c < 6$.}
\label{tab:PL2}
\end{table}

\section{Trace anomaly and equation of state}
\label{sec:trace_anomaly}

Power-like $1/T^2$ terms have also been found in the equation of state
of gluodynamics and
QCD~\cite{Pisarski:2006yk,Cheng:2007jq,Megias:2009mp}. The natural
observable to display such effects is the trace anomaly, or
interaction measure:
\begin{equation}
\epsilon - 3P = T^5\frac{d}{dT} \left(\frac{P}{T^4} \right) =
 \frac{\beta(g)}{2g} \langle (G_{\mu\nu}^a)^2 \rangle
\,.
\end{equation}
Unlike the pressure, the trace anomaly gets no contribution from the
ideal gas part and this is the primary quantity used in lattice to
obtain the pressure \cite{Boyd:1996bx}. Perturbatively, the quantity
$\langle (G_{\mu\nu}^a)^2 \rangle$ is proportional to $T^4$ up to
radiative corrections. Within the same model used to analyze the
Polyakov loop, it has a further power-like contribution proportional to
$T^2$ from the dimension 2 condensate~\cite{Megias:2009mp}:
\begin{equation}
\langle (G_{\mu\nu}^a)^2 \rangle^{\rm NP}  = 
-6 m_D^2 \langle A_{0,a}^2\rangle^\NP
,
\end{equation}
where $m_D$ is the Debye mass. The trace anomaly can then be expressed
in the following form
\begin{equation}
  \frac{(\epsilon - 3P)}{T^4} =\frac{(\epsilon - 3P)_{\rm pert}}{T^4} 
+ b_{\Delta} \left(\frac{T_c}{T} \right)^2 \,, 
\label{eq:e3pfitb}
\end{equation}
where $b_{\Delta}=-3 g \beta(g)\langle A_{0,a}^2\rangle^\NP/T_c^2$ for
three colors and no quarks. This pattern is similar to the one in
Eq.~(\ref{eq:Vqqpert}) for the $q\bar{q}$ potential and in
Eq.~(\ref{eq:PLpnp}) for the Polyakov loop.

One can make the same analysis that was performed in previous
sections, and fit lattice data for the trace anomaly from
Ref.~\cite{Boyd:1996bx} using Eq.~(\ref{eq:e3pfitb}). The weak
coupling expansion for the free energy is known up to order
$g_s^6\log(g_s)$~\cite{Shuryak:1977ut,Chin:1978gj,Kapusta:1979fh,Toimela:1982hv,Arnold:1994eb,Zhai:1995ac,Braaten:1995cm,Kajantie:2002wa}.
A renormalization group invariant (RGI) resummation, to be used below,
is presented in \cite{Megias:2009mp}. The perturbative series is
poorly convergent for the lattice QCD available temperatures $T < 5
T_c$, and even at much higher temperatures. There have also been
numerous attempts to resum perturbation theory in order to get a
better convergence of the result, one of the most developed techniques
being the hard thermal loop (HTL) perturbation theory. The free energy
of the gluon plasma has been computed recently up to three-loop order
in HTL~(see~\cite{Andersen:2009tc} and references therein).

In the fit of the lattice data for the trace anomaly we consider both
resummations (RGI and HTL), which enter as the term $(\epsilon -
3P)_{\rm pert}$ in Eq.~(\ref{eq:e3pfitb}). All the fits were performed
in a regime in which $\chi^2/{\rm dof} < 1$, so that reliable errors
could be extracted.  In particular, because PT is expected to work
better as the temperature increases, it is sufficient to change the lowest
temperature value of the interval. As fitting parameters we take
$b_\Delta$ and the parameter $\delta$ defined by $\mu = e^\delta 2\pi
T$. Because the running coupling depends on $\mu/\Lambda_{\rm QCD}$, a
change in $\delta$ is related to a change in $\Lambda_{\rm QCD}$, and
the correlation between $b_\Delta$ and $\Lambda_{\rm QCD}$ is measured
by the quantity $r(b_\Delta,\delta)$.

\begin{table}[htb]
\begin{center}
\begin{tabular}{|c|c|c|c|c|c|}
  \hline
  {\bf Order}  &  $\Delta_{\rm pert}/T^4|_{T=4.5 T_c}$ & $\delta$  &    $b_\Delta$   &   $r(b_\Delta,\delta)$ & $\chi^2/{\rm dof}$ \\
  \hline
  $\alpha_s$=const          & $-0.02(4)$  & $-$  & 3.46(13) & $-0.730$  & 0.35         \\
  ${\cal O}(\alpha^2)$    & $0.04^{+0.04}_{-0.02}$  & $2.3 \pm 1.9$  & 3.29(13)  & 0.722  & 0.86        \\
  ${\cal O}(\alpha^{5/2})$ & $-0.04^{+0.02}_{-0.06}$  & $2.3 \pm 1.6$  & 3.57(18) & $-0.867$  &  0.42        \\
  ${\cal O}(\alpha^{3})$   & $-0.07^{+0.03}_{-0.04}$  & $2.3 \pm 0.7$  & 3.73(19)  & $-0.880$   & 0.78         \\
  ${\cal O}(\alpha^{7/2})$ & $0.06^{+0.05}_{-0.02}$ & $2.3 \pm 0.9$ & 2.74(58)  & 0.891  & 0.99         \\
  ${\cal O}(\alpha^{4})$   & $-0.006 \pm 0.043$ & $0.45 \pm 2.4$  & 3.35(49) & 0.983  & 0.37    \\
\hline
\end{tabular}
\end{center}
\caption{Fit to trace anomaly lattice data from Ref.~\cite{Boyd:1996bx},
  $N_\sigma^3 \times N_\tau = 32^3 \times 8$, using
  Eq.~(\ref{eq:e3pfitb}). The temperature interval $ 1.13 < T/T_c <
  4.54$ is used for all lines except ${\cal O}(\alpha^{7/2})$, for which
  $2.0 < T/T_c < 4.54$ is taken. In the first line no running in
  $\alpha_s$ is applied. In the other lines the renormalization group
  invariant $\alpha(T) = 4\pi/(22 \log ( 2\pi T/\Lambda_{\rm QCD}))$ is
  used \cite{Megias:2009mp}. At ${\cal O}(\alpha^4)$ the value $A_6 =
  20.0$ is adopted \cite{Megias:2009mp}. In the first line, the
  correlation coefficient refers actually to $-r(b_\Delta,\alpha_s)$.
}
\label{tab:anomaly1}
\end{table}

\begin{figure}[tbp]
\begin{center}
\epsfig{figure=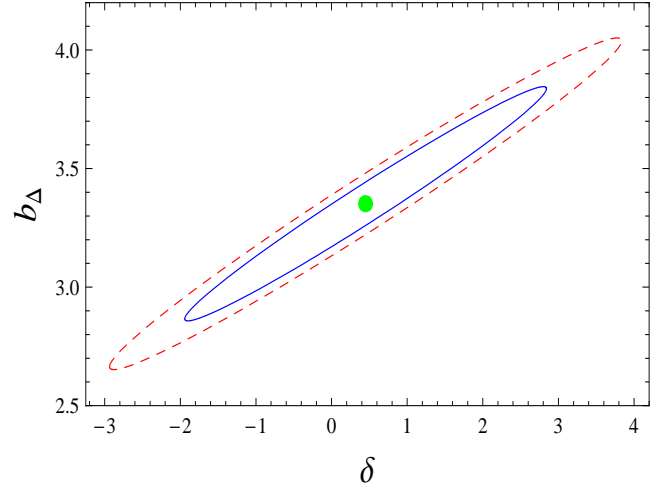,height=6.5cm,width=8.5cm}
\end{center}
\caption{Correlation ellipses corresponding to $\Delta \chi^2 = 2.3,
  4.6$ (${\rm dof}=8$) between the perturbative parameter $\delta$ and
  the non perturbative parameter $b_\Delta$ for the trace anomaly (see
  main text). The perturbative series $(\epsilon-3P)_{\rm pert}$ is
  considered to ${\cal O}(\alpha^4)$ in the RGI resummation of
  Ref.~\cite{Megias:2009mp}. The fit uses lattice data in the interval
  $1.13 T_c \le T \le 4.54 T_c$.}
\label{fig:Anomaly1}
\end{figure}

In the RGI resummation, at order ${\cal O}(\alpha^4)$ an undetermined
parameter, $A_6$, appears due to infrared divergences. Setting
$\delta=0$, we fit the lattice data using $b_\Delta$ and $A_6$ as free
parameters at order ${\cal O}(\alpha^4)$.  In this case, the best fit
in the regime $1.13 < T/T_c < 4.54$ gives~\cite{Megias:2009mp}:
\begin{equation}
  b_\Delta = 3.18(74)\,, \quad A_6 = 20.0 \pm 10.5  \,, 
\quad r(b_\Delta,A_6) = 0.992 \,, 
\label{eq:fite3pA6}
\end{equation}
with $\chi^2/{\rm dof} = 0.40$. Next we fit $b_\Delta$ and $\delta$
from ${\cal O}(\alpha^2)$ to ${\cal O}(\alpha^4)$. For the highest
order the central value of $A_6$ previously obtained is used. The
results are shown in Table~\ref{tab:anomaly1} and Fig.
\ref{fig:Anomaly1}.  The fit favors vanishing small perturbative
contributions, producing large values of $e^\delta$ $(> 10^3)$. To
prevent this, from ${\cal O}(\alpha^2)$ to ${\cal O}(\alpha^{7/2})$ an
upper bound $e^\delta<10$ has been set. Although the results are
not fully conclusive, in general, the correlation between perturbative and
non perturbative terms becomes larger when higher orders in PT are
included, and also, the value of the non perturbative coefficient
$b_\Delta$ tends to be smaller for the higher orders. The correlation
ellipses for ${\cal O}(\alpha^4)$ are displayed in
Fig.~\ref{fig:Anomaly1}.

We have also performed the analyses of this section using lattice data
from Refs.~\cite{Okamoto:1999hi} and \cite{Umeda:2008bd}. The values
of the parameters agree within estimated errors with those quoted here
from Ref.~\cite{Boyd:1996bx}. As a rule these alternative lattice data
lead to larger errors. For instance, the fit using the lattice data
from Ref.~\cite{Okamoto:1999hi} in the regime $1.14<T/T_c<3.6$ yields
$b_\Delta = 3.57(54)$, $3.25(46)$, $3.72(90)$, $3.73(80)$, $3.09(45)$,
$2.7(1.6)$, for orders from LO to ${\cal O}(\alpha^4)$ respectively.
The decrease of $b_\Delta$ as the PT order in increased is more
evident with these data, although they are affected by larger errors.
The correlation increases in this case from
$r(b_\Delta,\alpha_s)=-0.741$ at LO to $r(b_\Delta,\delta)=0.975$ at
${\cal O}(\alpha^4)$.

\begin{figure}[bbb]
\begin{center}
\vskip1cm 
\epsfig{figure= 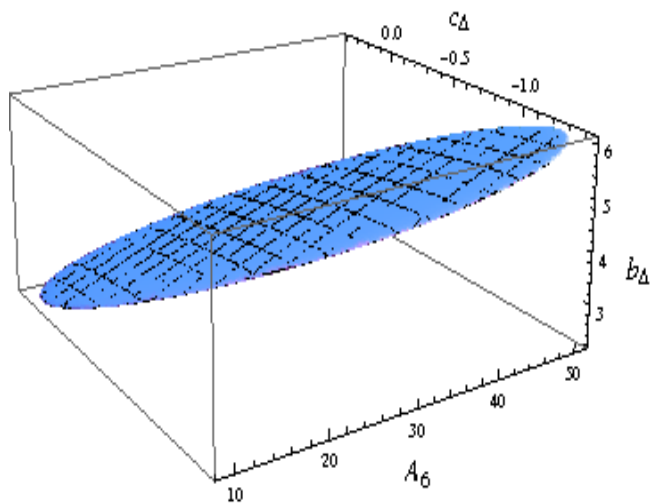,height=6.5cm,width=8.5cm}
\end{center}
\caption{Correlation ellipsoid corresponding to $\Delta \chi^2 = 3.53$
  (${\rm dof}=7$ and confidence level $68\%$) between the perturbative
  parameter $A_6$ and the non perturbative parameters $b_\Delta$ and
  $c_\Delta$ for the trace anomaly (see main text). The central values
  $A_6 = 29.9$, $b_\Delta = 4.15$ and $c_\Delta = -0.53$ follow from a
  fit in the interval $1.13 T_c \le T \le 4.54 T_c$.}
\label{fig:Anomaly3}
\end{figure}

In order to extract the possible contribution of a dimension four
condensate, we have also considered a fit using the formula:
\begin{equation}
  \frac{(\epsilon - 3P)}{T^4} = \frac{(\epsilon - 3P)_{\rm pert}}{T^4} + b_{\Delta} \left(\frac{T_c}{T} \right)^2 + c_\Delta \left( \frac{T_c}{T} \right)^2 \,. 
\label{eq:e3pfitbc}
\end{equation}
The results are shown in Table~\ref{tab:anomaly2}.  The parameter
$c_\Delta$ is compatible with zero, but a high correlation between
both non perturbative terms $b_\Delta$ and $c_\Delta$ is displayed, in
line with results of Sec.~\ref{sec:Polyakov_loop}. The correlation
$r(b_\Delta,\delta)$ tends to increase with the perturbative order, as
in Table~\ref{tab:anomaly1}.  The correlation ellipsoid corresponding
to a joint fit of $A_6$, $b_\Delta$ and $c_\Delta$ is displayed in
Fig.~\ref{fig:Anomaly3}.

\begin{table}[htb]
\begin{center}
\begin{tabular}{|c|c|c|c|c|c|c|}
\hline
{\bf Order}  & $\delta$  &    $b_\Delta$ & $c_\Delta$  &   $r(b_\Delta,\delta)$ & $r(b_\Delta,c_\Delta)$ & $\chi^2/{\rm dof}$ \\
\hline
$\alpha_s$=const             & $-$  & 3.7(7) &  $-0.3(9)$ & $-0.850$ & $-0.969$  & 0.31         \\
${\cal O}(\alpha^2)$       & $2.3 \pm 3.5$  & 3.09(51)  & $0.33 \pm 0.67$ & 0.747 & $-0.938$  & 0.80        \\
${\cal O}(\alpha^{5/2})$   & $1.6 \pm 3.1$  & $3.9 \pm 1.2$ & $-0.5 \pm 1.1$ & $-0.950$ & $-0.961$ &  0.30        \\
${\cal O}(\alpha^{3})$     & $2.3 \pm 2.1$  & $4.1 \pm 1.1$ & $-0.6 \pm 1.0$ & $-0.940$ & $-0.956$   & 0.31         \\
${\cal O}(\alpha^{7/2})$   & $2.3 \pm 1.9$  & 2.6(9)  & $1.1 \pm 1.1$ & 0.883 & $-0.926$ & 0.76         \\
${\cal O}(\alpha^{4})$     & $-0.6 \pm 1.2$  & $2.8 \pm 2.0$ & $-1.2 \pm 5.5$  & 0.983 & 0.960 & 0.29     \\
\hline
\end{tabular}
\end{center}
\caption{Same as Table~\ref{tab:anomaly2} but adding a 
  term $c_\Delta(T_c/T)^4$, c.f. Eq.~(\ref{eq:e3pfitbc}). 
  The interval $1.13 < T/T_c < 4.54$ has been used, except at
  ${\cal O}(\alpha^{7/2})$ where $1.24 < T/T_c < 4.54$ has been used.
  At ${\cal O}(\alpha^4)$ the value $A_6 = 20.0 \pm
  10.5$ has been adopted, cf. Eq.~(\ref{eq:fite3pA6}).}
\label{tab:anomaly2}
\end{table}

The analogous analysis can be carried out considering for
$(\epsilon - 3P)_{\rm pert}$ the HTL perturbation theory result at
1-loop~\cite{Andersen:1999sf}, 2-loops~\cite{Andersen:2002ey} and
3-loops~\cite{Andersen:2009tc}. The results are presented in
Tables~\ref{tab:anomalyHTL1} and \ref{tab:anomalyHTL2}.
\begin{table}[htb]
\begin{center}
\begin{tabular}{|c|c|c|c|c|c|}
  \hline
  {\bf Order}  &  $\Delta_{\mbox{\rm \tiny HTL}}/T^4|_{T=4.5 T_c}$    & $\delta$  &    $b_\Delta$   &   $r(b_\Delta,\delta)$ & $\chi^2/{\rm dof}$ \\
  \hline
  1-loop    & $-0.03^{+0.03}_{-0.04}$   & $-0.03 \pm 0.69$  & 3.69(40)  & 0.975  & 0.36        \\
  2-loops   & $-0.004^{+0.05}_{-0.004}$  & $-0.42 \pm 0.48$  & 3.57(28) & 0.949  &  0.41       \\
  3-loops   & 0.08(5)  & $2.3 \pm 9.7$  & $2.3 \pm 1.5$  & 0.977   & 0.67         \\
\hline
\end{tabular}
\end{center}
\caption{
  Fit of trace anomaly lattice data from Ref.~\cite{Boyd:1996bx},
  $N_\sigma^3 \times N_\tau = 32^3 \times 8$, using
  Eq.~(\ref{eq:e3pfitb}), in which $(\epsilon - 3P)_{\rm pert}$ is
  identified with the HTL
  result~\cite{Andersen:1999sf,Andersen:2002ey,Andersen:2009tc}. The
  interval $ 1.13 < T/T_c < 4.54$ is used for 1-loop and 2-loop orders,
  and $2.29 < T/T_c < 4.54$ at 3-loops.
}
\label{tab:anomalyHTL1}
\end{table}

\begin{table}[htb]
\begin{center}
\begin{tabular}{|c|c|c|c|c|c|c|}
\hline
{\bf Order}  & $\delta$  &    $b_\Delta$ & $c_\Delta$  &   $r(b_\Delta,\delta)$ & $r(b_\Delta,c_\Delta)$ & $\chi^2/{\rm dof}$ \\
\hline
1-loop    & $0.7 \pm 1.8$  & $4.3 \pm 1.0$  & $-0.6 \pm 1.1$ & 0.932  & $-0.976$  & 0.29        \\
2-loops  & $-0.47(80)$  & 3.58(40) & $-0.1 \pm 1.0$ & 0.411 & $-0.145$ &  0.47        \\
3-loops   & $2.3 \pm 9.3$  & $2.37 \pm 1.6$  & $1.3 \pm 1.3$ & 0.964 & $-0.898$   & 0.92         \\
\hline
\end{tabular}
\end{center}
\caption{Same as Table~\ref{tab:anomalyHTL1} but adding a term 
  $c_\Delta(T_c/T^4)$,
  c.f. Eq.~(\ref{eq:e3pfitbc}). The interval
  $1.13 T_c < T < 4.54 T_c$ is used at 1- and 2-loop orders, and
  $1.24 T_c < T < 4.54 T_c $ at 3-loops.}
\label{tab:anomalyHTL2}
\end{table}

A large correlation between $b_\Delta$ and $\delta$ (or $\Lambda_{\rm
  QCD}$) is found even at 1-loop order. In addition, the effect of a
smaller contribution from $b_\Delta$ at increasing perturbative orders
is rather clear in the HTL scheme. Once again, the correlation between
dimension 2 and dimension 4 condensates turns out to be
strong.\footnote{The relative small correlation at 2-loops in
  Table~\ref{tab:anomalyHTL2} is confirmed when other lattice data are
  used~\cite{Okamoto:1999hi,Umeda:2008bd}, but this result seems to be
  anomalous in view of Table~\ref{tab:anomalyHTL1}. Nevertheless at
  this order a high correlation between $c_\Delta$ and
  $\delta$, $r(c_\Delta,\delta)=0.840$, is found.}

\section{Conclusions}
\label{sec:conclusions}

We have proposed to study statistical correlations between
perturbative series and power corrections when analyzing lattice QCD
results as a way to probe quantitatively the duality proposed by
Narison and Zakharov in Ref.~\cite{Narison:2009ag}. We have found that
the effects of this duality start feeling at leading order in
perturbation theory, because even at this order the correlations are
very strong. We have performed an analysis for the heavy $q\bar{q}$
potential at zero temperature, and for several observables in the
deconfined regime of thermal QCD, for which power corrections were
derived in previous works.  As a byproduct, we have addressed the
important question on the finding of inverse power corrections at
finite temperature above the deconfinement phase transition. Our
analysis dissolves the apparent contradiction between the real
existence of thermal $1/T^2$ power corrections in the lattice results
and the persistent failure of perturbation theory to a given finite
order to reproduce them; these two extremely disjoint scenarios are
actually complementary and strongly interrelated.

The present correlation study can be extended to other cases where the
condensate-perturbative duality might be expected.  A very extreme
situation corresponds to just use power corrections and no
perturbation theory at all. This could be considered the starting
point of the analysis, where some ``non perturbative'' physics would
be expected. Another extreme situation requires a complete knowledge
of perturbation theory to all orders, a certainly unrealistic
situation.  For the situation in-between note that to observe a
decreasing condensate for increasing orders in perturbation theory it
is not at all trivial as it depends on the behavior of the
perturbative expansion. On the other hand, {\it showing} that the
residual condensate actually vanishes when {\it all} terms in
perturbation theory are taken into account seems to us as difficult as
solving QCD exactly.

\medskip

We acknowledge useful correspondence with the authors of
Ref.~\cite{Brambilla:2009bi}.



\end{document}